\documentclass[12pt]{article}
\usepackage{amsmath,amssymb,epsfig}

\usepackage{color}
\input{colordvi.tex}
\def\unit{{\relax{\rm 1\kern-.26em I}}}
\def \InN{{\unit^n_{N}}}
\def\InNc{{\unit^n_{N_c}}}
\def\unitnp{{({\unit'}^{N_c-n}_{N_c})}}

\newcommand{\lsim}{\lesssim}
\newcommand{\gsim}{\gtrsim}

\makeatletter

\renewcommand\section{\@startsection {section}{1}{\z@}%
                                   {-3.5ex \@plus -1ex \@minus -.2ex}%
                                   {2.3ex \@plus.2ex}%
                                   {\normalfont\large\bfseries}}

\renewcommand\subsection{\@startsection{subsection}{2}{\z@}%
                                     {-3.25ex\@plus -1ex \@minus -.2ex}%
                                     {1.5ex \@plus .2ex}%
                                     {\normalfont\normalsize\bfseries}}

\newcount\hour \newcount\minute
\hour=\time \divide \hour by 60
\minute=\time
\def\now{%
\ifnum \hour<13
  \ifnum \hour=0 \advance \hour by 12 \number\hour:\else \number\hour:\fi%
     \ifnum \minute<10 0\fi%
     \number\minute%
\ A.M.%
\else \advance \hour by -12 \number\hour:%
  \ifnum \minute<10 0\fi%
  \number\minute%
  \ P.M.%
\fi%
}

\makeatother

\begin{document}

\baselineskip=18pt  
\numberwithin{equation}{section}  
\allowdisplaybreaks  



%
%


\thispagestyle{empty}

\vspace*{-2cm}
\begin{flushright}
\end{flushright}

\begin{flushright}
\end{flushright}

\begin{center}

\vspace{2.0cm}

\vspace{0.5cm}
{\bf\Large Light Gaugino Problem in Direct Gauge Mediation}\footnote{Talks given at PASCOS 2011 (University of Cambridge, July 3-8) and at SUSY breaking '11 (CERN, May 16-21).}
\vspace*{1.5cm}

{\bf
Yutaka Ookouchi\footnote{Based on works in collaboration with K. Hanaki, M. Ibe, Y. Nakai and C-S. Park \cite{HIOP, NakaiOokouchi}. }
} \\
\vspace*{0.5cm}

{\it Institute for the Physics and Mathematics of the Universe (IPMU),\\ University of Tokyo, Chiba, 277-8583, Japan}\\
\vspace{0.1cm}

\vspace*{0.5cm}

\end{center}

\vspace{1cm} \centerline{\bf Abstract} \vspace*{0.5cm}

It has been known that in a wide class of direct gauge mediation models, the gaugino masses vanish at leading order in SUSY breaking. Recently, this phenomenon is understood in connection with the global structure of vacua in O'Raifeartaigh-type models. We review recent developments on this topic.

\newpage
\setcounter{page}{1} 



\section{Introduction}

Since the first celebrated discovery of dynamical supersymmetry (SUSY) breaking models in vector-like theories \cite{IzawaYanagida,IT}, it had taken more than ten years to recognize the genericity of dynamical SUSY breaking in vector-like models. A new avenue was opened up by Intriligator, Seiberg and Shih (ISS) \cite{ISS}, who overcame difficulties in the model-building accepting the possibility that SUSY can be broken on a metastable vacuum. One of the striking characteristics of SUSY breaking by vector-like models is simple realizations in string theory as emphasized in \cite{HoriOoguri,BraneI,BraneII,BraneIII,KawanoOO,OO1}. After the breakthrough, there has been drastic progress on this avenue \cite{Ookouchi} (see \cite{ISreview,KOOreview} for reviews) and now the notion that metastability is inevitable is widely accepted.

A fascinating feature of accepting metastability is the flexibility of model building. As demonstrated in \cite{Sweet,KOO,DM,Terning,Strassler} and many other literatures (see references in \cite{ISreview,KOOreview}),
such flexibility makes it possible to construct simple and phenomenologically viable models and various ideas such as gauge mediation \cite{GMI} and conformal sequestering \cite{Schmaltz:2006qs} were revisited in light of the flexibility. Among them, one interesting fact worthy of emphasizing is the light gaugino problem. As initially pointed out in \cite{IzawaTobe} for direct gauge mediation and in \cite{SeqI} for semi-direct gauge mediation, in a wide class of gauge mediation models, the leading order contribution to the gaugino masses vanishes regardless of how the R-symmetry is broken. Anomalously light gauginos are problematic because relatively heavy sfermions induce a large correction to the Higgs mass, reintroducing a fine-tuning problem. One possible way out may be to take the messenger scale to be very close to the supersymmetry breaking scale so that the subleading corrections are to be in the same size as the leading contribution. However, as was studied in \cite{Yonekura}, such a model is severely constrained by the recent Tevatron bound on the sparticle masses and the mass bound on a light gravitino.

Recently, Komargodski and Shih shed light on the origin of the light gauginos. In \cite{KS}, they related the vanishing gaugino masses at leading order and global structure of the vacua in renormalizable theories, and showed, based on the study of generalized O'Raifeartaigh models, that the pseudomoduli space must have a tachyonic direction somewhere to generate sizable gaugino masses. This analysis opens up a new possibility to avoid the anomalously light gaugino problem. Namely, the leading order gaugino mass generally does not vanish if  supersymmetry is broken in uplifted metastable vacua. As we will show below, this idea was initially employed in \cite{KOO} and further discussed in \cite{GKK,AbelKhoze}.

A challenge in constructing a direct gauge mediation model is to avoid the Landau-pole problem. Once we embed the gauge group of the standard model into a flavor group of the dynamical sector, there appear many particles which transform under the standard model gauge group. These fields contribute to the beta functions of the gauge coupling constants and drive them to a Landau pole below the unification scale. In \cite{KOO} the Landau pole problem was circumvented by taking the messenger scale as an intermediate scale, which is one of great advantages of vector-like model. So, there are two hierarchal mass scales, one is the supersymmetry breaking scale and the other is the messenger scale. This structure seems generic for constructing an uplifted vacua \cite{GKK} which is essential for obtaining a sizable gaugino mass as mentioned above. Since the Landau-pole problem was circumvented, the model is reliable in an extremely wide range of energy scale from TeV scale to sub-Planckian scale. This allows us to make cosmological predictions at high energy \cite{HIOP}.

A feature shared by a class of ISS-variants is spontaneous symmetry breaking of global symmetry, especially breaking of $U(1)_B$. This $U(1)_B$ is a characteristic feature of vector-like $SU(N_c)$ gauge theories, so 
one can find the ``typical'' predictions of this type of theories by seeking the cosmological implications of broken $U(1)_B$, which is the main subject of section 5. In general, a Goldstone mode accompanied by the symmetry breaking can acquire a light mass by Planck-suppressed (or another high scale) corrections. This light particle is called a pseudo-Nambu-Goldstone boson. Existence of the light particle is significant in cosmology because its life-time is sufficiently long and it may dominate over the critical density of the universe by its oscillation, which is similar to the moduli problem \cite{AHKY}. 

The simplest way to avoid overclosing the universe is to gauge the global symmetry. 
A cosmic string produced by spontaneous breaking of the gauged symmetry gives us a fascinating possibility to probe the hidden sector by observing gravitational waves. Since $SO(N)/Sp(N)$ gauge theories do not have the baryonic symmetry, there may be a chance to distinguish these theories and $SU(N)$ theory among ISS-variants. As mentioned above, direct gauge mediation models tend to have a high messenger scale to avoid the Landau-pole problem. Combined with the fact that $U(1)_B$ is broken at the messenger scale in many ISS-variants, the tension of the cosmic string is also at a high scale. A high scale cosmic string can generate high-frequency strong gravitational waves from a cusp, which can be detected by future experiments such as Advanced LIGO, LISA and BBO. As is often the case, any solitons made before the inflation can be diluted during the inflation. In this case, the scenario of avoiding the overclosure by gauging the global symmetry does not leave any signature. On the other hand, if the symmetry breaking scale is smaller than the Hubble parameter during the inflation, the cosmic string can survive and give rise to a signal which can be detected. According to the $7$-year WMAP data \cite{7years}, the Hubble parameter during the inflation has an upper bound of order $10^{14}$ [GeV], so if the actual inflation scale is not so small, our argument is applicable for a wide range of intermediate scale without diluting the signature.

The outline of the paper is as follows. We first review the light gaugino problem in direct gauge mediation. We will then review an argument by Komargodski and Shih \cite{KS} which show a connection of light gaugino with a global structure of potential. Then, following \cite{NakaiOokouchi}, we future explore the connection in more general situation where non-canonical Kahler potential contribute a stability of vacuum. We will then show an explicit dynamical SUSY breaking model \cite{KOO} in which the leading order of gaugino masses are nonzero, which give a existence proof of our general arguments. Finally we will present some cosmological implications in such models.

\section{Light gaugino problem}

In this section, we review some known facts on gaugino masses at leading order in supersymmetry breaking scale. If the SUSY breaking scale is much smaller than messenger masses, which is typical situation in direct gauge mediation as mentioned in the introduction, one can reliably use the technique of analytic continuation into superspace \cite{GR,AGLR}.

\subsection{Gaugino screening}

We first review anomalously small gaugino mass problem which have been observed quite frequently in direct gauge mediation models. Suppose a messenger sector having the following general superpotential interaction with supersymmetry breaking field $\langle X \rangle =M + \theta^2 F$,
\begin{eqnarray}
W=\sum_{ab} {\cal M}(X)_{ab} \phi^a \tilde{\phi}^b .
\label{Messenger}
\end{eqnarray}
where the messenger mass matrix $\mathcal{M}_{ab}(X)$ is a holomorphic function of $X$. In this case, the gaugino masses\footnote{Throughout this paper, we focus on Majorana gaugino masses. Including standard model adjoint fields allow us to add Dirac mass terms as firstly pointed out in \cite{Fayet1}. } are generated by integrating out the messengers $\phi^a$, $\tilde{\phi}^b$. The formula is given \cite{GR,KOO} by
\begin{equation}
m_{ \lambda}=-{g^2_{\rm SM} \over 16 \pi^2} F{\partial \over \partial X} \log \det \mathcal{M}(X).
\label{GauginoMass}
\end{equation}
If $\det {\cal M}(X)$ is constant, the leading contribution to gauginos vanish.

Now we consider a more general setup in which the K\"ahler potential of messengers is non-canonical: it has a dependence on the supersymmetry breaking field $X$. Let us derive the gaugino mass formula utilizing analytic continuation into superspace \cite{AGLR}.  Suppose the theory has $N$ pair of messengers $\phi^a, \tilde{\phi}^a \,\, (a=1, \dots, N)$ which are fundamentals and anti-fundamentals of the standard model gauge interactions, respectively. A generic K\"ahler potential we consider is
\begin{equation}
K=\sum_a {Z}_a(X,X^{\dagger})(\phi^{a \dagger} e^{V_{\rm SM}^{(\phi)}}\phi^a + \tilde{\phi}^{a \dagger} e^{V_{\rm SM}^{(\tilde{\phi})}}\tilde{\phi}^a ), \label{KahlerA}
\end{equation}
where ${Z}_a(X,X^{\dagger})$ are some real functions of $X, X^{\dagger}$.
Finally, the superpotential is given by \eqref{Messenger}.

One can extract the gaugino masses generated by integrating out the messenger fields from the wave function renormalization gauge chiral superfield. One should, however, use the physical gauge coupling $R$ rather than the holomorphic one, since the holomorphic coupling is not invariant under field rescaling \cite{AGLR}. As pointed out in \cite{AGLR}, contributions from messenger interactions to the gaugino masses are suppressed by additional loop factors. Thus, a non-canonical K\"ahler potential cannot contribute to the leading order gaugino mass. To see this, one may write down the physical coupling below the messenger scale. For the sake of simplicity, we assume the fermion mass matrix of the messengers is constant: ${\cal M}(X)=m$, so $W=m \phi \tilde{\phi}$. The physical mass is defined using wavefunction renormalization ${ Z}_M$ of the messenger at the scale,
$${
\mu_m^2 ={|m|^2 \over { Z}_M(\mu_m)^2}.
}$$
Below this scale, the physical coupling is given by
$${
R(\mu)=R^{\prime}(\mu_0)+{b \over 16\pi^2} \log {\mu^2 \over \mu_0^2}+ {1 \over 16 \pi^2} \log {|m|^2 \over \mu_0^2 { Z}^{\prime}_M (\mu_0)^2}+{T_G \over 8\pi^2}\log {{\rm Re} S(\mu)\over {\rm Re} S^{\prime}(\mu_0)}-\sum_r {T_r \over 8\pi^2} \log { { Z}_r (\mu)\over { Z}_r^{\prime}(\mu_0) },
}$$
where $r$ runs all matter fields in the SSM, $\mu_0$ is the cut-off scale of the theory, and $b$ is a coefficient of beta function below the messenger scale. $S(\mu)$ is a holomorphic gauge coupling and primed quantities are the ones above the messenger scale. Here, we see that ${{ Z}_M(\mu_m)}$ dependence drops out at low energy. Thus, a non-canonical K\"ahler potential does not contribute to the leading order of gaugino mass. Moreover, we could have assumed a spurion dependence of K\"ahler potential at the cut-off scale. Plugging the definition of real coupling $R^{\prime}(\mu_0)$ at the cut-off scale, we see that ${ Z}^{\prime}_{M}(\mu_0)$ dependence also cancels out. Therefore, the leading order gaugino masses are not affected by spurion dependence of K\"ahler potential of messengers at all. However, if we impose a spurion dependence in $S^{\prime}(\mu_0)$, it definitely contributes. Although it is nothing but adding gaugino masses by hand at the cut-off scale, it is contained in a frame work of the gauge mediation \cite{GGM}, since it vanishes in turning off the gauge coupling of the SSM. Usually in calculable models, these contributions, if exist, are generated by a heavy messenger around the cut-off scale and small compared to the leading term.

\subsection{Next to leading order gaugino mass}

Since the sfermion masses generally arise at leading order, the vanishing gaugino masses at leading order implies that there is a hierarchy between gaugino and sfermion masses. One may consider the next to leading order gaugino masses to solve the hierarchy. There are several sources for non-vanishing gaugino masses at next to leading order:
\begin{itemize}

\item While the gaugino masses leading order in $F$ at one-loop are prohibited, there is no problem for having non-vanishing gaugino masses at higher order in $F$. Explicit calculations show that the next leading order contribution arises at ${\cal O} ({F^3/ M_{\rm mess}^5})$ \cite{IzawaTobe}. One might hope that the gaugino masses can be comparable to sfermion masses if $F/M_{\rm mess}^2 \sim 1$. However, these higher order corrections are suppressed by small numerical coefficients in known examples and not sufficient to solve the hierarchy. Also, there is a phenomenological constraint on such a low scale mediation model \cite{Yonekura}.

\item Another possibility is that the gaugino masses are generated in ${\cal O}(F)$, but at higher loop level. In this case, using wave-function renormalization technique \cite{AGLR}, one can explicitly show that the leading order gaugino mass in $F$ at two-loop also vanishes if one-loop contribution does. Thus, the leading contribution is generated at best from three-loop diagrams. This contribution includes additional loop factors, so should be suppressed compared to the leading order. 

\item As discussed in the previous subsection, there could be a contribution of a heavy messenger at the cut off scale or above, which would be of order ${\cal O}(F/M_{\rm heavy})$. This type of contribution is suppressed to the leading order soft masses by a factor of ${\mathcal O} (M_{\rm mess}/M_{\rm heavy})$ compared to the leading order soft masses.

\end{itemize}
In any case, the gaugino masses are suppressed compared to the leading order and so to sfermion masses. This, combined with the current experimental lower bound for gaugino masses, indicates that the scales of the sfermion masses should be much higher than that of electroweak symmetry breaking, giving rise to fine tuning for the Higgs mass via top-stop loops. This is in contrast to the fact that relatively heavy gauginos at the messenger scale does not cause any problem because the sfermion masses are driven to be in the same order as gaugino masses at a lower scale by standard model renormalization group effects.

\section{Generating leading order gaugino mass}

It has been observed that the gaugino masses vanish at leading order in a wide class of direct gauge mediation model, regardless of how R-symmetry is broken. Recently, Komargodski and Shih (KS) shed light on this curious feature and clarified that the pseudomoduli space cannot be locally stable everywhere in order to generate sizable gaugino masses \cite{KS}. Here we firstly review their argument, then extend to a model with non-canonical Kahler potential \cite{NakaiOokouchi}. See \cite{aze,Up1,Up2,Up3,Up4,added1,added2} for recent developments on related topic.

\subsection{Gaugino mass and stability of pseudomoduli space}

The starting point of their investigation is a general Wess-Zumino model with a canonical K\"ahler potential and a renormalizable superpotential. In canonical form, the superpotential can be written as \cite{Ray}
$${
W = F X  + {1\over2}(\lambda_{ab}X+m_{ab}){\phi}_a{\phi}_b+{1\over6}\lambda_{abc}{\phi}_a{\phi}_b{\phi}_c~.
}$$
In this case, at tree level supersymmetry is broken and $X$ is pseudomoduli direction\footnote{In general, pseudomoduli space exists in F-term SUSY breaking as was firstly pointed out \cite{Fayet2} and systematically studied in \cite{Ray} recently.  }. 
Suppose $\phi_a$ fields are charged under the SM gauge group $G_{SM}=SU(3)\times SU(2)\times U(1)$. From \eqref{GauginoMass}, leading order of gaugino masses vanish when $\det (\lambda X +m)$ is constant. So, to generate the leading order of gaugino mass, it cannot be constant. In this case, the determinant of $\lambda X+m$ must be a polynomial in $X$, 
$${\det(\lambda X+m)=\sum c_i(\lambda,m)X^i~.
}$$
Thus, there must be places in the complex $X$ plane where it vanishes. Consider the theory around some such point $X=X_0$, and let $v$ satisfy
$${
(\lambda X_0+  m)v=0~.
}$$
This corresponds to a massless fermion direction. The corresponding boson mode $\delta\phi_i=v_i$ must be a tachyon; Naively, since a diagonal component of boson mass matrix is zero, 
\begin{equation}
{\cal M}^2_B= \left(
  \begin{array}{cc}
  (\mathcal{M}_F^{\ast}\mathcal{M}_F)_{a\bar{b}}  &
\mathcal{F}^{\ast}_{ab} \\
  \mathcal{F}_{\bar{a}\bar{b}} & (\mathcal{M}_F\mathcal{M}_F^{\ast})_{\bar{a} {b}}
    \end{array}
 \right),
\end{equation}
where ${\cal F}_{ab}= F^* (\partial_{X} {{\cal M}_F})_{ab}$, non-zero off-diagonal component gives rise to a negative mass eigenvalue (For a precise argument, see \cite{KS}). Interestingly we have arrived at an argument for the inevitability of metastability to generate sizable gaugino masses. On the other hand, when the $\det (\lambda X+m)$ is constant, moduli space is stable everywhere in messenger direction. However, the leading order gaugino masses vanish. This clarifies a general symptom that many calculable direct gauge mediation model suffered from.

\subsection{Stability of messenger directions}

As we reviewed in the previous section, in every renormalizable O'Raifeartaigh-type
model, the pseudomoduli space cannot be stable everywhere to generate gaugino masses.
However, a renormalizable model is not always a good description of dynamical
SUSY breaking at low-energy. In many SUSY breaking models, correction terms in K\"ahler potential
are not negligible. Here, we will show that such terms
affect crucially the connection between gaugino masses and the landscape of vacua \cite{NakaiOokouchi}. 

Let us start with a general argument for the stability of
messenger directions. Suppose we have a superpotential interaction, 
\begin{equation}
W={\cal M}_F (X)_{ab}{\phi}^a\tilde{\phi}^b+F X,
\end{equation}
where $X$ is a chiral superfield which is responsible of SUSY breaking and $\phi, \tilde{\phi}$ are messengers. The lower indices of the messenger mass matrix ${\cal M}_F$ denote the derivatives with respect to messenger fields. When we turn on a generic non-canonical K\"ahler potential, $X$ direction is not
necessarily pseudo-flat as discussed in \cite{Ray}. Nevertheless, in order to focus on the stability of messenger directions at a point of the pseudomoduli space like \cite{KS}, we can keep a flat direction by imposing the following condition on the metric
\cite{AM},
\begin{equation}
\partial_X g^{X\bar{X}}\big|_{0} =0, \label{cond1}
\end{equation}
where $|_0$ denotes $\langle \phi^a  \rangle=\langle \tilde{\phi}^a  \rangle =0$. It is easy to check that the scalar potential with this condition keeps $X$ direction flat. 

In this setup, the boson mass-squared matrix of the messengers is given by 
\begin{equation}
{\cal M}_B^2=\left(
  \begin{array}{cc}
  (\mathcal{M}_F^{\ast}\mathcal{M}_F)_{a\bar{b}} - {\cal A}_{a\bar{b}} &
\mathcal{F}^{\ast}_{ab} \\
  \mathcal{F}_{\bar{a}\bar{b}} & (\mathcal{M}_F\mathcal{M}_F^{\ast})_{\bar{a} {b}}-
{\cal A}_{\bar{a}b}
  \end{array}
 \right).
\end{equation}
Here,
\begin{equation}
{\cal F}_{ab}= F^* (\partial_{X} {{\cal M}_F})_{ab},\quad {\cal
A}_{a\bar{b}}= {R_{a \bar{b}}}^{X\bar{X}} |F|^2,
\end{equation}
where ${R_{a\bar{b}}}^{X\bar{X}}$ are components of the Riemann tensor. We simply assumed
$g_{X\bar{X}}=1$ at $\tilde{\phi}^a=\phi^b=0$. Suppose $v_a$ is a unit vector satisfying $(\mathcal{M}_F)_{ab} v_b = 0$. Then, a bosonic
mode corresponding to this direction has a mass, 
\begin{equation}
\begin{split}
 &\left(
 v^\dagger~ v^T
 \right)
 {\cal M}_B^2
 \left(
  \begin{array}{c}
   v \\
   v^{\ast}
  \end{array}
 \right)= v^{T}\mathcal{F}v -v^\dagger {\cal A} v + c.c.
 \end{split}
\end{equation}
If ${\cal A}v=0$ or simply if ${\cal A}=0$, then the bosonic mode must be massless in order to have a consistent vacuum, or we have to allow the vacuum to have a tachyonic direction. However, in general, this does not true. As we will demonstrate below, one can easily lift a tachyonic direction and make the pseudomoduli space stable everywhere by using the contribution from the non-canonical part of K\"ahler potential ${\cal A}$.

With this in mind, in the rest of this section, we will try to construct the model which has non-zero leading order gaugino masses and a pseudomoduli space that is locally stable everywhere. For simplicity, let us focus on a specific model the model  with non-canonical K\"ahler potential. The superpotential is given by
\begin{equation}
W= \lambda X(\phi_1\tilde{\phi_1}+\phi_2\tilde{\phi_2}) + m\phi_1\tilde{\phi_2} + fX.
\end{equation}
This model with canonical K\"ahler potential has a tachyonic direction around $\langle X \rangle = 0$, so we will try to lift this direction by introducing non-canonical K\"ahler potential,
\begin{equation}
K=|X|^2+\left(1+ {|X|^2\over M^2} \right) \left(|\phi_1|^2+ |\tilde{\phi}_2|^2 \right)+\left(1- {|X|^2\over M^2} \right) \left(|\tilde{\phi}_1|^2+|\phi_2|^2 \right), \label{noncanonical}
\end{equation}
where $M$ is a large cut-off scale of the theory and we have required vanishing messenger mass supertrace so that our model is UV insensitive \cite{Poppitz,EGGM}. Since the above K\"ahler potential satisfies the condition \eqref{cond1} given in the previous subsection, the pseudo flat-direction of $X$ is kept. There is a zero eigenvalue in the fermion mass matrix at $\langle X \rangle = 0$, and here the eigenvalues of the boson mass-squared matrix of messengers are 
\begin{equation}
\frac{1}{2}\left(\,{m}^{2} \pm \,\sqrt {{m}^{4}+4\,{\lambda}^{2}{f}^{2}-4\,(f/M)^{2}{m}^{2}+4\,(f/M)^{4}} \right).
\end{equation}
We can impose a condition between parameters of the model such as ${\lambda}^{2}{f}^{2}-\,(f/M)^{2}{m}^{2}+\,(f/M)^{4}<0$ so as not to have any tachyonic direction. As discussed in section 2, since non-canonical K\"ahler potential of messengers does not contribute to the gaugino mass at the leading order, the leading order gaugino mass is given in the same fashion as the case with canonical K\"ahler potential,
\begin{equation}
m_{\tilde{g}} \sim \frac{f}{\langle X \rangle}.
\end{equation}
Here, the expectation value of $X$ can be estimated by stabilizing the one-loop effective potential. The Coleman-Weinberg potential in this kind of models has been calculated in \cite{AM}, which claims that $X$ does not stabilize at the origin and so R-symmetry is spontaneously broken even in the case with non-canonical K\"ahler potential. Therefore, we can obtain non-zero leading order gaugino masses in the model with a pseudomoduli space that is locally stable everywhere.

While we have considered a model with a pseudo-flat direction, as we have seen in the previous subsection, the existence of pseudomoduli is not guaranteed in models with non-canonical K\"ahler potential. So in the next subsection, we will show a model which generate leading order of gaugino mass on a global minimum.

\subsection{Sizable gaugino mass on the global minimum}

When we consider the case where there is no pseudomoduli space, it becomes unclear how we can generalize the statement of the Komargodski-Shih's argument. We are interested in a connection between the leading order gaugino masses and metastability of the vacuum. Then, we will try to solve the question whether we can obtain non-vanishing gaugino masses on the global minimum or not. The answer is yes. In \cite{Nomura:1997ur}, the authors obtained non-vanishing leading order gaugino masses on the global minimum. However, they used a dynamical SUSY breaking model and the resulting model is incalculable. Then, for our current purpose, we do not need to focus on dynamical SUSY breaking, so we can take our familiar O'Raifeartaigh-type model.

The explicit model of the SUSY breaking sector is a $U(1)$ gauge theory whose superpotential is given by
\begin{equation}
W = X_0(f+\lambda \varphi_1\varphi_2)+ m(X_1\varphi_1+X_2\varphi_2).
\end{equation}
The $U(1)$ charge assignments of $X_0$, $X_1$, $X_2$, $\varphi_1$ and $\varphi_2$ are $0$, $-1$, $1$, $1$ and $-1$ respectively.
We call this $U(1)$ gauge interaction as the messenger gauge interaction.
We can take all couplings, $\lambda, m, f$ as real without loss of generality and assume $f \ll m^2$. On the SUSY breaking vacuum, $\langle X_1 \rangle = \langle X_2 \rangle = \langle \varphi_1 \rangle = \langle \varphi_2 \rangle = 0$ and $X_0$ has a non-zero F-term.

Next, consider the messenger sector. The simplest possibility for our purpose would be the following, 
\begin{equation}
W_{mess}= y_q S q\tilde{q} +y_ES E\tilde{E} + {\kappa \over 3} S^3,
\end{equation}
where $q$ and $\tilde{q}$ are messengers charged under the standard model gauge symmetries and $S, E, \tilde{E}$ are the standard model gauge singlets. Only $E, \tilde{E}$ have charges $1,-1$ under the messenger $U(1)$ gauge interaction. We also take couplings $y_q, y_E, \kappa$ as real. When we integrate out the SUSY breaking sector, two-loop correction generates positive scalar masses for fields $E$ and $\tilde{E}$ like usual gauge mediation, which is given by
\begin{equation}
m_E^2 = m_{\tilde{E}}^2 \sim \left({g_{mess}^2\over 16\pi^2} \right)^2 \left({\lambda f \over m} \right)^2,
\end{equation}
where $g_{mess}$ is the coupling of the messenger gauge interaction. As pointed out in \cite{Nomura:1997ur}, these positive scalar masses generate negative mass squared by one-loop effects of $E$ and $\tilde{E}$ such as
\begin{equation}
-m_S^2 \simeq {4\over 16\pi^2}y^2_E m_E^2 \ln {\Lambda \over m_E},
\label{Smass}
\end{equation}
where $\Lambda$ is the cut-off scale and we assume $y_E \lesssim 1$ so that $m_E^2 \gg |m_S^2|$ is satisfied.
Then, the effective scalar potential of the messenger sector including these corrections is given by
\begin{eqnarray}
V_{mess}&=&\big|{y_E S \tilde{E}} \big|^2+\big|y_E{SE} \big|^2+\big|{y_q S \tilde{q}} \big|^2+\big|y_q{Sq} \big|^2+\big|{y_E E\tilde{E}+y_q q\tilde{q}+\kappa S^2} \big|^2\nonumber \\
&&+m_E^2 |E|^2 +m_E^2 |\tilde{E}|^2+m_S^2 |S|^2. 
\end{eqnarray}
This potential is minimized at
\begin{equation}
\langle |S|^2 \rangle ={|m_S^2| \over 2\kappa^2},\ \  \langle q \rangle =\langle \tilde{q} \rangle =\langle E \rangle =\langle \tilde{E} \rangle =0.
\end{equation}
Note that the expectation value of the SUSY breaking field $S$ is uniquely determined and there is no pseudomoduli space in the messenger sector. The contribution to the vacuum energy is given by
\begin{equation}
V_0=-{m_S^4 \over 4 \kappa^2}.
\end{equation}
This vacuum is the global minimum in certain parameter range. The standard model gaugino mass can be calculated as
\begin{equation}
m_{\tilde{g}} \sim  {\langle |F_S| \rangle \over \langle S \rangle } = \frac{|m_S|}{\sqrt{2}}.
\end{equation}
Therefore, we obtain the leading order gaugino mass on the global minimum of the potential, unlike direct gauge mediation without additional gauge interactions.

\section{A dynamical model}

In previous section, we reviewed argument by Komargodski and Shih and showed inevitability of metastable vacuum for calculable direct gauge mediation model. The idea of higher energy vacuum to generate large gaugino mass is initially employed 
 in \cite{KOO}. Here we will show the model and demonstrate all ideas shown in previous section in dynamical model.

Let us remind you of the original ISS model \cite{ISS}, which is an $SU(N_c)$ gauge theory with $N_f$ flavors $Q_i$ and $\bar Q_i$ with a superpotential $W= \sum_{i=1}^{N_f} m_i Q^i \bar Q_i$.
We consider a case where the mass matrix $m_i$ is a diagonal with $N_f-N_c$ of its eigenvalues being $m_0$ and $N_c$ of them $\mu_0$: $m_i=\mathrm{diag}(m_0,\cdots,m_0,\mu_0,\cdots,\mu_0)$.
In this case, we may write the superpotential more explicitly as
\begin{eqnarray}
 W_{\rm mass} = m_0 (Q^I \bar Q_I) + \mu_0 (Q^a \bar Q_a)\ ,
\end{eqnarray}
where $I=1,\cdots, N\equiv N_f- N_c$ and $a=1,\cdots, N_c$ and  the color $SU(N_c)$ indices are contracted in $(Q \bar Q)$. To this superpotential we add the following superpotential
\begin{eqnarray}
W_{\rm def}= - \frac{1}{m_X} (Q^I \bar Q_a) (Q^a \bar Q_I)\ ,
\label{eq:higher-d}
\end{eqnarray}
The interaction term \eqref{eq:higher-d} breaks the $U(1)_R$ symmetry.
Hence the remaining global symmetry is $SU(N) \times SU(N_c) \times U(1)_{P} \times U(1)_B$. In the magnetic description, the theory is described by the meson fields
\begin{equation}
Y^I_{~J} = Q^I \bar Q_J \;,\qquad Z^I_{~a}= Q^I \bar Q_a\;,\qquad \tilde Z^a_{~I} = Q^a \bar Q_I \;,\qquad \Phi^a_{~b} = Q^a \bar Q_b\;.
\end{equation}
The superpotential in the magnetic description is given by
\begin{equation}\label{KOOSUP}
W= h \mathrm{Tr} \left[ m^2 Y + \mu^2 \Phi - \chi Y \tilde{\chi} - \chi Z\tilde{\rho} - \rho \tilde{Z}\tilde{\chi} - \rho \Phi \tilde{\rho} - m_z Z \tilde Z\right],
\end{equation}
where $q$ and $\chi$ are components of magnetic quarks, and $ m^2 \equiv m_0  \Lambda$ and $\mu^2 \equiv \mu_0 \Lambda$. By redefining fields with appropriate phase rotations, we may restrict ourselves to the case of real $m$, $\mu$ and $m_z$ without loss of generality.

As long as the deformation superpotential \eqref{eq:higher-d} is small, the meta-stable vacuum identified in the ISS model still exists.
However this deformation introduces additional meta-stable vacua far away from the origin.
These vacua have lower energy densities than the ISS vacuum and are given by
\begin{equation}\label{E:nVacConfig}
\begin{split}
&Y^I_{~J} = \frac{\mu^2}{m_z} \left(\InN\right)^I_J\;,\qquad \Phi^a_{~b} = \frac{m^2}{m_z} \left(\InNc\right)^a_{~b}+\gamma_* \unitnp^a_{~b}\\
&\chi^I_{~J} = m \delta^I_J\;,\qquad \tilde\chi^I_{~J} = m \delta^I_J\\
&\rho^I_{~a} = \mu \Gamma^I_{~a}\;,\qquad \tilde\rho^a_{~I} = \mu \Gamma^a_{~I}\\
&Z^I_{~a} = -\frac{m\mu}{m_z}\Gamma^I_{~a}\;,\qquad \tilde Z^a_{~I} = - \frac{m\mu}{m_z}\Gamma^a_{~I}\;,
\end{split}
\end{equation}
$n$ can be any integer between 0 and $N$.
$\Gamma^a_{~I}$ and $\Gamma^I_{~a}$ have 1 in the first $n$ diagonal elements and 0 elsewhere.
That is,
\begin{equation}
\Gamma^a_{~I} = \begin{pmatrix} \unit_n & 0_{n\times ({N-n})} \\ 0_{(N_c-n)\times n} & 0_{(N_c-n)\times (N-n)} \end{pmatrix}\;,\qquad
\Gamma^I_{~a} = \begin{pmatrix} \unit_n & 0_{n\times (N_c-n)} \\ 0_{(N-n)\times n} & 0_{(N-n)\times (N_c-n)} \end{pmatrix}\;.
\end{equation}
$\unit_m$ is an $m\times m$ identity matrix and $\unit^p_q$ is a $q\times q$ matrix whose first $p$ diagonal elements are 1 and 0 otherwise. As studied in \cite{HIOP}, all pseudo-moduli are stabilized by the Coleman-Weinberg potential. $\gamma_*$ is one of the stabilized moduli.

The symmetry preserved in a generic supersymmetry breaking vacuum is $SU(n) \times SU(N_f - N_c - n) \times SU(N_c -n) \times U(1)^2$. By gauging $SU(N-n)$, we can easily get the leading order of gaugino mass,
$${
m_{\lambda}\simeq -{g^2 {(N_c-n)}\over (4\pi)^2} {h \mu^2 m_z \over  m^2}\left( 1 +{\cal O}({m_z^2  \over m^2}) \right).
}$$
Similarly, when we gauge the $SU(N_c-n)$ group, gaugino masses are given by
$${
m_{\lambda}= -{g^2 {(N-n)}\over (4\pi)^2} {h \mu^2 m_z \over  m^2}\left( 1  +{\cal O}({m_z^2  \over m^2})\right).
}$$
However, if we gauge the $SU(n)$ symmetry, since the messenger does not include a tachyonic direction anywhere in the moduli space, one expects that gaugino masses vanish according to \cite{KS}. By a direct computation of mass matrix it is easy to show ${\rm det}M=\mu^2 m_z$. Clearly the determinant of the matrix is $X$ independent, so gaugino masses vanish.

\section{Cosmological aspects}

Now we have understood a new avenue for model building of direct gauge mediation. It would be interesting to explore cosmological implication in light of this avenue. Here we will review work \cite{HIOP}. 

\subsection{General argument}

First of all, we will describe the models we are interested in and discuss their various features shared by many of the models. Specifically, we will consider metastable SUSY breaking in vector-like models.
One striking recent discovery is that metastable SUSY breaking is generic in vector-like models.
Various phenomenologically viable models have been constructed based on metastable vacua, and so it would be interesting to also explore cosmological implications common to these models. Actually, as we will emphasize below, direct-type models are highly constrained from observation and also predictive, and there are several features common in a class of such models. Typically, energy scales in the theories tend to be high and that makes it easy for those models to be tested cosmologically. Another interesting feature is the presence of Nambu-Goldstone boson, which will play a crucial role in the following arguments.

A particular feature worth emphasizing in our setup is a connection between the mass scale of messengers and that for the $U(1)_B$ symmetry breaking. In general, these two scales have nothing to do with each other. In the direct gauge mediation models exploiting ISS-variants, however, the two scales coincide, which enables us to probe messengers by gravitational effects through cosmic strings.
We suppose the hidden sector has two hierarchal mass scales, supersymmetry breaking scale $\mu$ and messenger scale $m$. These two parameters are related by the condition that soft supersymmetry breaking parameter in the supersymmetric standard model (SSM) should be of ${\cal O}(\rm TeV)$, or equivalently, 
\begin{equation}
{\mu^2 \over m}\sim 100 \ {\rm [TeV]}. \label{CD1}
\end{equation}
We suppose that its SUSY breaking effect is mediated to SSM by gauge mediation and that the hidden sector is in an uplifted vacuum to avoid having light gauginos. As discussed in previous sections, in O'Raifeartaigh-like model, to get non-vanishing leading order gaugino mass, the existence of an uplifted vacuum is crucial. Moreover, as emphasized in \cite{GKK}, for the existence of a metastable uplifted vacuum, two scales are necessary. For flavor violating Planck scale physics effects to be sufficiently small, we also assume that the supersymmetry breaking scale satisfies $\mu \lsim 10^{9.5}$ [GeV]. As for the messenger scale $m$,
we assume that SSM are reliable up to sub-Planckian scale (i.e., there is no Landau pole up to sub-Planckian scale) and take it to be an intermediate scale. Especially, we are interested in a range
\begin{equation}
m^{\rm expt}_{\rm obs}\le m \leq H_{\rm inf} \label{mrange}\;,
\end{equation}
The lower bound is coming from the condition that the gravitational waves can be accessible in future observation and will be explained in more details below. The upper bound is the condition that the Hubble parameter $H_{\rm inf}$ during the inflation is larger than the messenger scale. In this case, the cosmic strings generated from the breaking of certain gauge symmetries at the scale $m$ in our models are not inflated away
According to the WMAP $7$-year data \cite{7years}, an upper bound of the inflation scale is $H_{\rm inf}\lsim 1.6\times 10^{14}$ [GeV]. To make our scenario the most interesting, we assume that the parameter is around $H_{\rm inf} \sim 10^{14}$ [GeV] and assume no secondary inflation below the scale.

\subsection{PNGB mass and decay width}

As we emphasized in the introduction, $U(1)_B$ symmetry in a class of ISS-variants makes a distinct feature: if it is a global symmetry, its spontaneous breaking generates a Nambu-Goldstone boson. On the other hand, as is well-known, any global symmetry should be broken in string theory. So, it should be explicitly broken at the Planck scale if it is not gauged. Here we will show that in our assumption, explicit breaking of the symmetry causes cosmological disaster in a wide range of parameter space. 

If there is no particular reason, the superpotential should have the following form of symmetry breaking term in the electric theory:
$${
W_{\rm ele} \supset   {1\over M_{\rm pl}^{N_c-3}}Q^{N_c},
}$$
where we omitted the epsilon tensor. In the magnetic dual description, it is written by
$${
W_{\rm mag}\supset {\Lambda^{2N_c-N_f} \over  M_{\rm pl}^{N_c-3}}q^{N_f-N_c} .
}$$
where dual quark can be written as $q={(\chi, \rho )}$ in term of the notation of the previous section. The dynamical scale $\Lambda$ of $SU(N_c)$ gauge theory has been incorporated due to the dimensional analysis. As studied in \cite{HIOP}, dominant contribution to the mass of the pseudo-Nambu-Goldstone boson is given by
$${
(m_{\rm PNGB})^2 = {\Lambda^{2N_c-N_f+3} \over M_{\rm pl}^{N_c}} m^{N_f-N_c-1}.
}$$
With this mass of the pseudo-Nambu-Goldstone boson, let us compute its decay probability and life time. Because of CP and C symmetries, decay processes are highly surpressed by Planck or messenger scales. Thus, the inverse of the decay probability, $\tau_{\rm PNGB} \sim \Gamma^{-1}$, is larger than the current age of the universe $\tau_0\sim 4\times 10^{17}\, {\rm [s]}$ \cite{HIOP}.

\subsection{Overclosing the universe}

Now we are ready to study energy density of an oscillation of pseudo-Nambu-Goldstone boson. 
After an inflation, if the radiation-dominated era starts quickly, the inflaton decay must be the most efficient and the reheating temperature becomes of order $ 10^{16} {\rm [GeV]}$. However, the large reheating temperature generically makes it hard to solve the gravitino problem \cite{MMY}. To make reheating temperature low, the inflaton decay has to be inefficient, so there should exist inflaton oscillating era. Hence we consider the following two cases. 

Firstly, let us discuss the case where the reheating temperature is lower than the one when PNGB starts oscillating\footnote{In this paper we denote the Planck mass $M_{\rm pl}\simeq 1.9\times 10^{19}$ GeV and the reduced Planck mass $M_{r.p.}\simeq 2.4\times 10^{18}$ GeV.}, $T_{\rm PNGB}\simeq \sqrt{M_{r.p.} m_{\rm PNGB}}$. At the beginning the energy density of the universe is dominated by inflaton field, $\rho_{\rm I}$. At some point, inflation ends and oscillation of inflaton era starts. In this era, enegy density of the inflaton deceases as $a^{-3}$. When the Hubble parameter becomes equal to the mass of PNGB (point C in figure 1), energy density of the inflaton is given by $\rho_I^{(C)}=m_{\rm PNGB}^2 M_{\rm r.p.}^2$. Finally inflaton decays and the universe is reheated (see point D in the figure 1). There, since the radiation dominate ear begins. So energy density of inflaton is written in terms of the reheating temperature as $ \rho^{(D)}_I =T^4_R$. 

On the other hand, energy density of the PNGB at a point A is just given by potential energy $\rho^{(A)}_{PNGB}=m_{\rm PNGB}^2 m^2$. Since the ratio 
${\ln \rho_{PNGB}^{(B)} / \ln \rho_{PNGB}^{(A)}}$ should be equal to ${\ln \rho_I^{(D)} / \ln \rho_I^{(C)}}$, one get the energy density at the beginning of the radiation-dominated era, $\rho^{(B)}_{PNGB}=m^2 T_R^4/ M_{\rm r.p.}^2$. In the radiation dominated era, energy density of scale as $T^4$ while PNGB is still oscillating,
\begin{eqnarray}
\rho_{PNGB}(T)=\rho^{(B)}_{PNGB} \left({T\over T_R} \right)^3 \\
\rho_{I}(T)=\rho^{(D)}_{I} \left({T\over T_R} \right)^4 .
\end{eqnarray}
Therefore, these two densities coincide at some point. The temperature at that point $T=T_{dom}$ is given by 
$${
T_{dom}=T_R {m^2 \over M_{\rm r.p.}^2 }
}$$
If this temperature were comparable or larger than $1$[eV], the energy density of PNGB oscillation would also be comparable or larger than that of the standard model matters and would overclose the universe. One can avoid the overclosure by taking the reheating temperature to be very low. This leads to the condition
$${T_R < 10\  {\rm [GeV]} \qquad {\rm when}\quad m=10^{13} {\rm [GeV]}. 
},$$
which is is logically possible, but it require unnaturally-tuned low reheating temperature.

\begin{figure}[htbp]
\begin{center}
  \epsfxsize=7cm \epsfbox{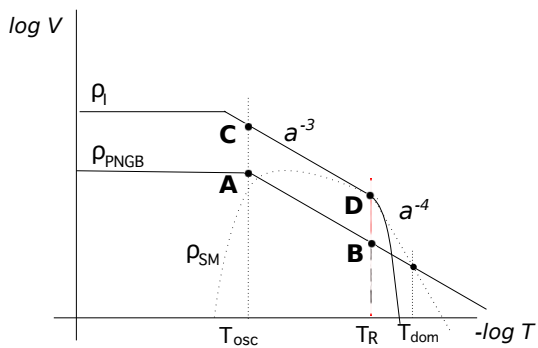}
\end{center}
 \vspace{-.5cm}
 \label{aafirstcase}
\caption{\sl \small The evolution of the density as the universe cools down. The two solid lines represent the density for inflatons and that for pseudo-Nambu-Goldstone bosons, respectivly, and the dotted line represents that for Standard Model matters. To avoid the overclosure of the universe, the line for $\rho_{PNGB}$ and that for $\rho_{SM}$ should intersect below $T=1[eV]$.}
\end{figure}

Another possible scenario is the case where $T_{\rm PNGB}< T_R$. Doing the same analysis as above, one can estimate the temperature when PNGB oscillation dominates the energy of the universe, $T_{dom}^{\prime}=T_{\rm PNGB} {m^2 / M_{\rm r.p.}^2 }$. To take this temperature to be lower than $1$[eV], however, one needs require the mass of pseudo-Nambu-Goldstone boson to be very small ($m_{\rm PNGB} < 10^{-20}$[GeV]), which is undesirable if one tries to avoid the Landau pole.

We, therefore, conclude that any of the scenarios discussed in this section generally leads to a cosmological disaster. 
A simple wayout is to gauge the $U(1)_B$ symmetry and it will in turn lead to a cosmologically interesting possibility. This gauging eliminates the existence of PNGB, but instead dynamically generates cosmic strings with tension of order the symmetry breaking scale $m$. 
As we will discuss below, however, the tension of the cosmic strings is constrained by the observation of gravitational waves.
So, we are led to a highly exciting hypothesis that the parameters in the hidden sector can be determined by astrophysical observation: the messenger scale is determined by the detection of gravitational waves, while the supersymmetry breaking scale is fixed by data and the fixed messenger scale. In the model \cite{KOO}, there are two types of string defects. one is the Abrikosov-Nielsen-Olesen type and the other is semi-local.

The most relevant question to topological defects is when it is formed. If the Hubble parameter during the last inflation $H_{\rm inf}$ exceeds the symmetry breaking scale $m$,
or alternatively, if the highest temperature\footnote{
The actual highest temperature $T_{\rm H}$ that a system can reach is given by $T_{\rm H} \simeq (T_{\rm R}^2 H_{\rm inf} M_{\rm pl})^{1/4}$ where $T_{\rm R}$ is the standard reheating temperature. 
} of the breaking sector
after inflation $T_{\rm H}$ exceeds $m$, cosmic strings are formed \cite{domainI,domainII}
$${
{\rm max}\, [H_{\rm inf},T_{\rm H}] \gsim  m.
}$$
It is known that if we assume only one inflation in the history and do not assume another origin of entropy production, reheating temperature is highly constrained by the gravitino problem \cite{MMY} in gauge mediation scenario. An upper bound of reheating temperature is determined roughly by the SUSY breaking scale that is much smaller than the messenger scale in our setup. So most particles in the SUSY breaking sector are not thermalized including cosmic strings, and $H_{\rm inf}> T_{\rm H}$ in our scenario. If $H_{\rm inf}$ is greater than $m$ then the transition of $U(1)_B$ symmetry may well not be completed during inflation, and cosmic string is formed when the Hubble parameter $H_{\rm inf}$ is of order $m$ by Kibble-Zurek mechanism. However, if $H_{\rm inf} < m$,
then the cosmic string will be inflated away.

As pointed out by Vilenkin \cite{VilenkinGW}, since the energy of a cosmic string turns into gravitational waves, there should be a stochastic gravitational wave background. The wave length of a gravitational wave originating from loops is typically comparable to the size $L$ of the emitting string loop. The frequency is red-shifted due to a subsequent cosmic expansion, so the frequency we observe today, $\omega$, is smaller than that
at the production
\begin{equation}
\omega ={L^{-1} \over 1+z_{\rm red}}.
\end{equation}
Observations of pulsar timing give an upper bound on $G\mu_T$ for ultra-low frequency $f\sim 10^{-9,-8}$ [Hz]. As of 2008 it is of order $G\mu_T <{\cal O}(10^{-8})$. Although pulsar timing is only for ultra-low frequency range, future experiments such as LISA and LIGO will give us data for complementary bands of frequency. LISA will observe a gravitational waves in frequency range $10^{-4}\sim 10^{-1}$ [Hz]. LIGO and Adv LIGO (LIGO II) is around $10\sim 10^{4}$ [Hz]. The strongest future experiment is  BBO, which probes around $10^{-2}\sim 10^{2}$ [Hz]. Since these observations probe mutually different frequency ranges, they can cover a wide range of frequency and can exclude the existence of cosmic strings of order $G\mu_T >{\cal O}(10^{-8})$. 

On the other hand, if we focus on a specific frequency, one can get finer data. For example, LIGO can access $G\mu_T \sim {\cal O} (10^{-11})$ around $150$ [Hz]. Also, LISA at $3.88\times 10^{-3}$ [Hz] can probe $G\mu_T \sim {\cal O}(10^{-13})$. If these experiments detect gravitational wave around this scale, it is indeed fascinating to us because we can probe our hidden sector by using gravitational waves! If not, that would give us one of the nice stringent constraints for model building of gauge mediation.

At high frequency range, there is a fascinating possibility of observation. A reconnection of strings generates a cusp. A large energy is concentrated at the tip and it emits an intense beam of gravitational waves. The Fourier transform of a cusp singularity is much larger at high frequency. So there is a big chance to detect it in future experiments.

\section*{Acknowledgments}

YO's research is supported by World Premier International Research Center Initiative (WPI Initiative), MEXT, Japan.

%
%

\end{document}